\documentclass[journal,transmag]{IEEEtran}

\newcommand{\comments}[1]{}

\ifCLASSINFOpdf
\else
\fi
\usepackage{amsmath,latexsym,amssymb}
\usepackage{ctable}
\usepackage{multirow,multicol}
\usepackage{subcaption}
\usepackage{graphicx}
\usepackage{mdwlist}
\usepackage{wrapfig}
\usepackage[utf8]{inputenc}
\usepackage[T1]{fontenc}
\usepackage{hyperref}
\usepackage{ctable}
\usepackage{moredefs}
\usepackage{changes}
\usepackage[light,condensed,math]{kurier}

\renewcommand{\comments}[1]{}

\begin{document}
%
\title{Locally Linear Embedding and fMRI feature selection in psychiatric classification}


\author{\IEEEauthorblockN{Gagan Sidhu\IEEEauthorrefmark{1}
\IEEEauthorblockA{\IEEEauthorrefmark{1}Department of Computing Science, 1-337 Athabasca Hall,
University of Alberta, Edmonton, AB T6G 2E8}
Corresponding author: Gagan Sidhu (email: gagan@g-a.ca).}}

%



\maketitle
\begin{abstract}
\hspace{-1.10em}Background:\\
Functional magnetic resonance imaging (fMRI) provides non-invasive measures of neuronal activity using an endogenous Blood Oxygenation-Level Dependent (BOLD) contrast. This article introduces a nonlinear dimensionality reduction (Locally Linear Embedding) to extract informative measures of the underlying neuronal activity from BOLD time-series. The method is validated using the Leave-One-Out-Cross-Validation (LOOCV) accuracy of classifying psychiatric diagnoses using resting-state and task-related fMRI.\\
Methods:\\
Locally Linear Embedding of BOLD time-series (into each voxel's respective tensor) was used to optimise feature selection. This uses Gau\ss' Principle of Least Constraint to conserve quantities over both space and time. This conservation was assessed using LOOCV to greedily select time points in an incremental fashion on training data that was categorised in terms of psychiatric diagnoses.\\
Findings:\\
The embedded fMRI gave highly diagnostic performances (> 80\%) on eleven publicly-available datasets containing healthy controls and patients with either Schizophrenia, Attention-Deficit Hyperactivity Disorder (ADHD), or Autism Spectrum Disorder (ASD). Furthermore, unlike the original fMRI data before or after using Principal Component Analysis (PCA) for artefact reduction, the embedded fMRI furnished significantly better than chance classification (defined as the majority class proportion) on ten of eleven datasets.\\
Interpretation:\\
Locally Linear Embedding appears to be a useful feature \comments{\replaced{extraction}{selection}}extraction procedure that retains important information about patterns of brain activity distinguishing among psychiatric cohorts.
\end{abstract}



\IEEEdisplaynontitleabstractindextext

%
\IEEEpeerreviewmaketitle

\section{Introduction}
%
%
%
%
\IEEEPARstart{O}{ver} a century ago, Charles Darwin alluded to an experimental paradigm that involved direct observation of the brain's physical mechanisms (nervous matter)~\cite{darwin1871descent}, where such observations~\cite{roysher} would serve as the physical basis for dichotomising species-specific behaviour.\comments{according to the stimulus adminstered at time \textit{t}. The main caveat with Darwin's approach is the assumption that the physical manifestations of mental activity, presumably reflected in the movements of the brain's nervous matter, can be measured.} In the early 90s, an indirect and non-invasive measurement of mental activity over uniformly-spaced time points became possible through functional Magnetic Resonance Imaging (fMRI), which allows paramagnetic deoxyhemoglobin to act as an endogenous Blood Oxygenation-Level Dependent (BOLD) contrast~\cite{ogawa3}. This article pursues Darwin's proposed experimental paradigm by Locally Linear Embedding (LLE)~\cite{lle1} the BOLD time-series to produce precise summaries of cerebral activity that may optimise the classification of different brain states, such as mental disorders.
\comments{measured BOLD response (over time) in a topology~\cite{bourbaki} defined over Cartesian space with the Pythagorean distance metric, which is expected to produce precise measurements of mental activity that will assist in dichotomising healthy animals from those with mental disorders.}

\par Logothetis \textit{et al} found that an increase in invasively-measured neural activity directly and monotonically reflects local BOLD signal increases and, for short stimulus presentations, there is a linear relationship between BOLD and neural responses~\cite{nikos2}. This suggests the unobservable neural activity is spatially-localised in anatomical space. Locally Linear Embedding of fMRI data in space and time can, in principle~\cite{archibald1914,compkern,vsepr} (see Appendix~\ref{app:LLE}), summarise these \textit{local} measurements of neuronal mass activity~\cite{natmrirev}--via the notion of \textit{analytic capacity}~\cite{melnikov}--to disclose information about \textit{global} (i.e., whole-brain) activity patterns.

\section{Methods: Testing Locally Linear Embedding (LLE) with Cross Validation}
\par The discriminatory power of Locally Linear Embedding was compared to the original fMRI, both before and after applying Principal Component Analysis (PCA)~\cite{pcabook} for artefact reduction~\cite{nikos2}. This comparison was performed on eleven datasets containing cohorts with different mental disorders, using a combination of Leave One Out Cross Validation (LOOCV) on the training set, with greedy feature selection based on Fisher discriminability~\cite{fisher1,fisher2}. The purpose of using LOOCV and feature selection is to find the time points that discriminate patients from controls on the respective dataset. The feature selection step initially starts with an empty candidate set of time points and proceeds to select the time point with the highest discriminatory power~\cite{sfs}. Then, time points that improve discrimination in conjunction with those already in the candidate set are added to this set in incremental fashion; this process is terminated when there are no more time points that can be added to the candidate set to improve discriminability. Note that the selection of time points is based upon cross validation and does not induce any biased sampling.

\par To illustrate the patterns that best discriminate between groups, a paired two-sample t-test between the patient and control groups is performed to both threshold and identify the statistically-significant differences (\textit{p} $<$ 0.05 uncorrected) in space (at the time points identified by the greedy feature selection). Per Mill's Methods of Induction (Method of difference), the functional differences (depicted by the statistically significantly-different regions) at the respective time point are therefore a necessary part of the cause of the phenomena that distinguish the subject groups~\cite{sol1}, which in this case pertain to a neuropsychiatric disorder. Neuropsychiatric disorders are diagnosed using clinical assessments that include: evaluating the background demographics, collecting first and third party observations, and a structured psychiatric interview with the subject~\cite{psychassess}. In detail:

\comments{removed by dr friston: This process reflects the intuition that differences in mental activity between patients and controls can occur, and therefore be identified, at any time point during the scan, where these time points can then be objectively thresholded to identify statistically significant physical differences. Per Mill's Methods of Induction (Method of difference), the functional differences (depicted by the statistically significantly-different regions) at the respective time point are therefore a necessary part of the cause of the phenomena that distinguish the subject groups~\cite{sol1}, which in this case pertain to a neuropsychiatric disorder. Given the simplicity of the feature selection and that cohorts' resting-state fMRI are not synchronised -- i.e., cohorts are not administered identical stimuli at the same time during the scan-- the predictive power of the diagnostic volumes determined on resting-state data is anticipated to vary. Given that differences in mental activity can occur at any time, it follows that the number of diagnostic time points can vary between datasets with the same experimental design and subject groups, especially for resting-state data.}

%

Every subject's fMRI time-series is treated as a four-dimensional array $\mathbf{X}\in\mathbb{R}^{  L \times W \times H \times T}$ with $V=LWH$ $T$-dimensional voxel waveforms $\mathbf{x}_i \in \mathbb{R}^T$ for $i=1,\ldots,V$. Assume each subject scan $\mathbf{X}_i$ is associated with a binary-valued class label $y_i$ representing the diagnosis and that, for any subject scan, every voxel waveform $\mathbf{x} \in \mathbb{R}^T$ is generated by a vector $\mathbf{z}\in\mathbb{R}^d$ corresponding to a point on the manifold. Our approach to fMRI-based diagnosis involves two stages:

\noindent \textbf{fMRI reconstruction} takes the subject's fMRI as input and outputs a reconstructed fMRI that is more informative than the original. Formally, this reconstruction is a mapping $\mathcal{L}: \mathbb{R}^{L\times W\times H \times T} \to \mathbb{R}^{L\times W \times H \times d}.$ 

In other words, Locally Linear Embedding reduces a time-series of length \textit{T} to a smaller number of spatial modes of dimensionality \textit{d}; these modes contain all the information used for the subsequent step.\\
\textbf{Classification} builds a classifier that takes the subject's reconstructed fMRI as input, and outputs a class label $y_i\in\{0,1\}$. The classifier is therefore a mapping $\mathcal{C}: \mathbb{R}^{L\times W\times H \times d} \to \{0,1\}$.

The reconstructed, or reduced, fMRI data produced from step 1 is hereon referred to as $\mathbf{Z}$. All reconstructions initially vectorise the fMRI data to produce a two-dimensional array $\mathbf{X}\in\mathbb{R}^{V\times T}$, and conclude by reshaping the resulting two-dimensional reconstruction $\mathbf{Z}\in\mathbb{R}^{V \times d}$ into a four-dimensional array $\mathbf{Z}\in\mathbb{R}^{L\times W \times H \times d}$.

\paragraph{Principal Component Analysis (PCA)~\cite{bishop,pcabook}} reconstructs $\mathbf{X}$ by finding an orthogonal rotation that minimises the reconstruction cost

\begin{equation*}\label{eqn:pca}
\min_{\mathbf{B} =[\mathbf{b}_1,\ldots,\mathbf{b}_T]} \sum_{i=1}^V || \mathbf{x}_{i} - \mathbf{\bar{x}} - \mathbf{B}\mathbf{z}_i||_2^2
\end{equation*}

\noindent where $\bar{\mathbf{x}} \in \mathbb{R}^T$ is the mean over all voxel waveforms, and $\mathbf{z}_i = \mathbf{B}^\intercal(\mathbf{x}_i-\mathbf{\bar{x}}) \in \mathbb{R}^T$. To find the optimal $\mathbf{B}$, compute the right-hand matrix for the singular value decomposition (SVD) of $\mathbf{X} = \mathbf{UDB}^\intercal$, which contains the $T$ right-singular vectors of $\mathbf{X}$~\cite{htf}. It follows that $\mathbf{Z} = \mathbf{XB} \in \mathbb{R}^{V \times T}$ is the rotated matrix that minimises the reconstruction cost of the subject's fMRI, where every column of $\mathbf{Z}$ is a \textit{principal component}. It is assumed the first $d$ principal components (columns) of $\mathbf{Z}$ capture the ``systematic structure", where the confounding factors are relegated to the remaining \textit{T} - \textit{d} principal components, which produces the two-dimensional reconstructed fMRI $\mathbf{Z}\in\mathbb{R}^{V\times d}$.

\par Using PCA's ``systematic structure" for distinguishing humans with different neurological disorders has been met with caution~\cite{classcalhoun}, largely because PCA's application to fMRI has some subjective components~\cite{pcafmri}. The main limitation behind the PCA reconstruction is that it assumes the lower-dimensional manifold is a linear subspace.\comments{inability to measure the \textit{local} spatiotemporal covariance between voxels, which which requires a graphical model to define the spatial neighbours of every voxel.} We demonstrate that introducing the Cauchy stress tensor~\cite{cauchy} on the Cartesian space with the Pythagorean distance metric enables three-dimensional measurements over time, thereby revealing the local (group) action in the physical system.

\paragraph{Locally Linear Embedding (LLE)~\cite{lle1,LLE2}} reconstructs $\mathbf{X}$ by constructing the Cauchy stress tensor~\cite{cauchy} at every voxel $i$ for $i=1,\ldots,V$, which is achieved by minimising the reconstruction cost of its waveform $\mathbf{x}_i$ in terms of its spatially-adjacent neighbours:	
\begin{equation} \label{eqn:objfunc}
\min || \mathbf{x}_{i} - \sum_{j \in \mathcal{N}(i)} w_{i,j}\mathbf{x}_{j}||_2^2, \;\;\;\;\; \textrm{where} \;\; \sum_{j \in {\cal N}(i)} w_{i,j} = 1
\end{equation}

\noindent where the neighbourhood set ${\cal N}(i)$ for voxel $i$ is the complement of its $K$ spatial neighbours on the surface of the sphere with radius \textit{r}, and $\mathbf{w}_i = [w_{i,1},\ldots,w_{i,|{\cal N}(i)|}] \in \mathbb{R}^{|{\cal N}(i)|}$ are the reconstruction weights containing the spatially-invariant geometric properties of the Cauchy stress tensor at voxel \textit{i}.

To determine the \comments{second-order invariant properties of the Cauchy stress tensor}\textit{analytic capacity}~\cite{melnikov} at every voxel location \textit{i}, LLE first subtracts the \textit{K} spatial patterns of the voxels on the boundary of the sphere centred around voxel \textit{i} to determine the separation distance from the origin of the tensor at the respective voxel. Then, it computes the local (symmetric) spatiotemporal covariance matrix:
\begin{equation}
\begin{split} 
\mathbf{G}_i &= \mathbf{C}_i^\intercal\mathbf{C}_i\\ &= [(\mathbf{x}_j-\mathbf{x}_i),...,(\mathbf{x}_{j+|\mathcal{N}(i)|} - \mathbf{x}_i)]^\intercal  [(\mathbf{x}_j - \mathbf{x}_i),..., (\mathbf{x}_{j+|\mathcal{N}(i)|} - \mathbf{x}_i)] \\&+ \xi \mathbf{I}_{|{\mathcal{N}}(i)|}
\end{split}
\end{equation}
where\comments{ that conditions the $\mathbf{C}_i$ is a Hilbert space.} $\xi\mathbf{I}_{|{\cal N}(i)|}$ is a non-negative regularisation term to enforce positive-definiteness (for this study, $\xi=0$). LLE calculates the reconstruction weights by finding the unique minimum-norm solution~\cite{pinv} to the constrained least-squares problem defined by:
\begin{align}
\mathbf{G}_i \mathbf{w}_i = \mathbf{1}_{|{\cal N}(i)|} &\iff \mathbf{G}_i^+  \mathbf{1}_{|{\cal N}(i)|} = \mathbf{w}_i
\end{align}
\noindent where $\mathbf{1}_{|{\cal N}(i)|} \in \mathbb{R}^{|{\cal N}(i)|}$ is a vector of ones and the $j^{th}$ element of $\mathbf{w}_{i,j}$ can be thought of as the average height of a curve representing the \textit{mean transit time of the indicator}~\cite{lasseningvar} of voxel $j$ from voxel $i$ over the duration of the scan. Since $\mathbf{G}_i$ represents the squared distance of the surface forces from voxel $i$, the reconstruction weights $\mathbf{w}_i$ are Lebesgue measures~\cite{lebesgue} summarising the \textit{analytic capacity}, or spatially-invariant geometry~\cite{gregory}, of the space-filling curve~\cite{hilbert}, where the constraint ensures that the areas between the imaginary surface (acting as the origin that divides the body) and curves (defined by the stress vectors) are 1 in each of the $|\mathcal{N}(i)|$ directions. In practice the weights can be brittle~\cite{mlle} due to any number of reasons. Modified Locally Linear Embedding (MLLE) therefore computes the $1 \leq s_i \leq \textit{K}$ linearly-independent (orthogonal) vectors\footnote{When using MLLE it is possible for $d > K$, thus the optimal number of weight vectors $s_i$ for each voxel $i$ is determined by setting $d=1$ so that \textit{K - 1} $\leq s_i \leq K$--i.e., $s_i$ is set to span as large of a basis as possible. After this step the desired dimensionality \textit{d} is then input to the eigensolver.} $\mathbf{Q}_i \in\mathbb{R}^{K \times K}$ of $\mathbf{G}_i$ using the eigendecomposition $\mathbf{G}_i = \mathbf{Q}_i^\intercal\mathbf{A}\mathbf{Q}_i$, thereby allowing the definition of multiple weight vectors for each voxel. Assuming the columns (eigenvectors) $[\mathbf{q}_1,\ldots,\mathbf{q}_K] = \mathbf{Q}_i$ are sorted in descending order of their respective eigenvalues $\lambda_1^{(i)},\ldots,\lambda_K^{(i)}$, MLLE uses the first $s_i$ columns to compute multiple local weight vectors for a single voxel:
\begin{multline*}
 \mathbf{w}_i^{(\ell)} = (1-\alpha_i)\mathbf{w}_i + \mathbf{Q}_i\mathbf{H}_i^{(\ell)}\;\;\textrm{where:}\;\; \alpha_i = \frac{1}{\sqrt{s_i}}||\mathbf{Q}_i^\intercal \mathbf{1}_K ||_2^2\\
 \mathbf{H}_i^{(\ell)}\in\mathbb{R}^{s_i},\;\;\mathbf{H}_i = \mathbf{I} - 2\mathbf{h}\mathbf{h}^\intercal\in\mathbb{R}^{s_i\times s_i},\;\;\mathbf{h} \in \mathbb{R}^{s_i},\;\textrm{and}\\ s_i = \max_{\ell}\Bigg\{\ell \leq K- d, \frac{\sum_{p=K-\ell+1}^K\lambda_p^{(i)}}{\sum_{p=1}^{K-\ell}\lambda_p^{(i)}} < \eta \Bigg\}
\end{multline*}
\noindent where $\mathbf{h} = \frac{\mathbf{h}_0}{||\mathbf{h}_0||}$ if $\mathbf{h}_0 = \alpha_i\mathbf{1}_{s_i} - \mathbf{Q}_i^\intercal\mathbf{1}_K \neq \mathbf{0}\in\mathbb{R}^{s_i}$ (else $\mathbf{h} = \mathbf{h}_0 = \mathbf{0}\in\mathbb{R}^{s_i}$), $\eta = \boldsymbol\rho_{\lceil V/2 \rceil}$, $\boldsymbol{\hat\rho}_i = \frac{\sum_{p=d+1}^K\lambda_p^{(i)}}{\sum_{p=1}^d \lambda_p^{(i)}}$, $\boldsymbol\rho = \mathrm{sort}(\boldsymbol{\hat\rho},\mathrm{ascending})$, and $\boldsymbol{\hat\rho},\boldsymbol\rho \in\mathbb{R}^{V}$.
Since every measure's invariant properties are determined in a square-integrable space~\cite{schmidt}, LLE performs a global least-squares optimisation based on Gau{\ss}' Principle of Least Constraint~\cite{gauss} to calculate the vectors $\mathbf{z}_1,\ldots, \mathbf{z}_V$ corresponding to points on the manifold:
\begin{multline}\label{eqn:llered}
\mathbb{E}[\mathbf{Z}] = \min_{ \mathbf{Z} = [\mathbf{z}_1,\ldots,\mathbf{z}_V]} \sum_{i=1}^V \sum_{\ell=1}^{s_i} || \mathbf{z}_{i} - \sum_{j \in \mathcal{N}(i)} w_{i,j}^{(\ell)}\mathbf{z}_{j}||_2^2, \;\;\;\\ \textrm{ such that} \;\;\; \mathbf{Z}\mathbf{Z}^\intercal = \mathbf{I}
\end{multline}
\noindent where $d \leq T$ is the dimensionality parameter selected by the user and $\mathbf{Z}\in\mathbb{R}^{V \times d}$ is the two-dimensional reconstructed fMRI. The global optimisation therefore calculates the points on the manifold~\cite{sherr,seunglee} that act as the four-dimensional orthogonal basis that best retains the geometry of the stress vectors. These represent the second order invariant properties of each voxel's Cauchy stress tensor. A detailed explanation of this optimisation is provided below.

\par Define $\mathbf{\hat{W}}_i \in \mathbb{R}^{V \times s_i}$ as the local sparse adjacency matrix, where:
\begin{multline*}
\mathbf{\hat{W}}_i(\mathcal{N}(i),:) = \mathbf{w}_i,\;\; \mathbf{\hat{W}}_i(i,:) = -\mathbf{1}_{s_i}^\intercal,\;\;\textrm{and} \\\mathbf{\hat{W}}_i(j,:) = 0,\;\; \forall j \not\in \{\mathcal{N}(i)\cup i \}
\end{multline*}
The optimisation in Equation~\ref{eqn:llered} can be written as a minimisation of the expected reconstruction cost, or error:
\begin{equation}
\mathbb{E}[\mathbf{Z}] = \sum_{i=1}^V ||\mathbf{Z}\hat{\mathbf{W}}_i||_2^2 = \mathrm{trace}(\mathbf{Z}\sum_{i=1}^V\hat{\mathbf{W}}_i\hat{\mathbf{W}}_i^\intercal\mathbf{Z}^\intercal) = \mathrm{trace}(\mathbf{Z}\boldsymbol\Phi\mathbf{Z}^\intercal)
\end{equation}
\noindent where $\mathbf{\hat{W}}_i\mathbf{\hat{W}}_i^\intercal$ is the orthogonal projection for voxel $i$, and $\boldsymbol\Phi =\sum_{i=1}^V\mathbf{\hat{W}}_i\mathbf{\hat{W}}_i^\intercal$ is the sparse, symmetric and positive-definite \textit{alignment matrix}, and therefore admits the eigendecomposition~\cite{drmackay}:

\begin{equation}
\mathbf{Z}\boldsymbol\Phi\mathbf{Z}^\intercal=\mathbf{Z}\boldsymbol{\Lambda}\mathbf{Z}^\intercal
\end{equation}
where $\boldsymbol{\Lambda} \in \mathbb{R}^{(d+1) \times (d+1)}$ is the diagonal matrix containing the ${d+1}$ smallest eigenvalues of $\mathbf{Z}\boldsymbol\Phi\mathbf{Z}^\intercal$, and $\mathbf{Z} \in \mathbb{R}^{V \times (d+1)}$ are the corresponding eigenvectors; Rayleigh's variational principle~\cite{rayleigh,courant} enables calculation of these bottom $(d+1)$ eigenvectors. Each eigenvector represents a degree of freedom in space and time, where the $(d+1)$ eigenvector is the global unit vector that fills three-dimensional space. The global unit vector is discarded to enforce the constraint that the manifolds have mean zero.

\noindent \textbf{Note:} To avoid degenerate solutions, LLE requires the manifolds to be centred around the origin in \textit{both} space and time -- i.e., $\sum_{i}^V \mathbf{Z}_{i,:} = \mathbf{0}\in\mathbb{R}^d$ and $\sum_{i}^d \mathbf{Z}_{:,i} = \mathbf{0} \in\mathbb{R}^{V}$--  \textit{and} also have outer products with unit covariance -- i.e., $\mathbf{Z}\mathbf{Z}^\intercal = \mathbf{I}$. Centring the manifolds about the origin ensures they are of the same scale, which is superficially similar to the common practice of signal, or count rate, normalisation~\cite{neuronuclear}. The unit covariance constraint imposes the requirement that the reconstruction errors of the extracted manifolds are measured on the same scale. 
%
%


\begin{table*}[t]
\tabcolsep 0.25pt
  {\fontsize{5.5}{7}\selectfont
  \centering
    \caption[Caption for LOF]{Results.}\label{tbl:results}    
    \begin{tabular}{rrrrrrrrrrrrrrrrrr}
    \toprule
    \multicolumn{1}{c}{\multirow{2}[-5]{*}{Dataset}} & \multicolumn{1}{c}{\multirow{2}[-5]{*}{Partition}} & \multicolumn{4}{c}{Specificity} & \multicolumn{4}{c}{Sensitivity} & \multicolumn{4}{c}{Precision} & \multicolumn{4}{c}{Accuracy} \\
    \midrule
    \multicolumn{1}{c}{} & \multicolumn{1}{c}{} & \multicolumn{1}{c}{Chance} & \multicolumn{1}{c}{Original} & \multicolumn{1}{c}{LLE} & \multicolumn{1}{c}{PCA} & \multicolumn{1}{c}{Chance} & \multicolumn{1}{c}{Original} & \multicolumn{1}{c}{LLE} & \multicolumn{1}{c}{PCA} & \multicolumn{1}{c}{Chance} & \multicolumn{1}{c}{Original} & \multicolumn{1}{c}{LLE} & \multicolumn{1}{c}{PCA} & \multicolumn{1}{c}{Chance} & \multicolumn{1}{c}{Original} & \multicolumn{1}{c}{LLE} & \multicolumn{1}{c}{PCA} \\
{Beijing} & Training & \multicolumn{1}{l}{100\%} & \multicolumn{1}{l}{73.3\% $\pm$ 15.8\%} & \multicolumn{1}{l}{80\%\;\,\, $\pm$ 14.3\%} & \multicolumn{1}{l}{66.7\% $\pm$ 16.9\%} & \multicolumn{1}{l}{0\%} & \multicolumn{1}{l}{53.3\% $\pm$ 17.9\%} & \multicolumn{1}{l}{93.3\% $\pm$ 8.9\%} & \multicolumn{1}{l}{80\% \;\;\,$\pm$ 14.3\%} & \multicolumn{1}{l}{0\%} & \multicolumn{1}{l}{66.7\% $\pm$ 16.9\%} & \multicolumn{1}{l}{82.4\% $\pm$ 13.6\%} & \multicolumn{1}{l}{70.6\% $\pm$ 16.3\%} & \multicolumn{1}{l}{50\% \;\;\,$\pm$ 17.9\%} & \multicolumn{1}{l}{63.3\% $\pm$ 17.2\%} & \multicolumn{1}{l}{\textbf{86.7\% $\pm$ 12.2\%}} & \multicolumn{1}{l}{73.3\% $\pm$ 15.8\%} \\
   (Peking\_3)       & Holdout & \multicolumn{1}{l}{100\%} & \multicolumn{1}{l}{75\%} & \multicolumn{1}{l}{87.5\%} & \multicolumn{1}{l}{25\%} & \multicolumn{1}{l}{0\%} & \multicolumn{1}{l}{100\%} & \multicolumn{1}{l}{50\%} & \multicolumn{1}{l}{25\%} & \multicolumn{1}{l}{0\%} & \multicolumn{1}{l}{66.7\%} & \multicolumn{1}{l}{66.7\%} & \multicolumn{1}{l}{14\%} & \multicolumn{1}{l}{67\%} & \multicolumn{1}{l}{83.3\%} & \multicolumn{1}{l}{\textbf{75\%}} & \multicolumn{1}{l}{25\%} \\
         \multirow{2}[0]{*}{COBRE} & Training & \multicolumn{1}{l}{100\%} & \multicolumn{1}{l}{57.7\% $\pm$ 13.6\%} & \multicolumn{1}{l}{92.3\% $\pm$ 7.3\%} & \multicolumn{1}{l}{96.2\% $\pm$ 5.3\%} & \multicolumn{1}{l}{0\%} & \multicolumn{1}{l}{60\%\;\;\, $\pm$ 13.4\%} & \multicolumn{1}{l}{88\% \;\;\,$\pm$ 8.9\%} & \multicolumn{1}{l}{52\%\;\;\, $\pm$ 13.7\%} & \multicolumn{1}{l}{0\%} & \multicolumn{1}{l}{57.7\% $\pm$ 13.6\%} & \multicolumn{1}{l}{91.7\% $\pm$ 7.6\%} & \multicolumn{1}{l}{92.9\% $\pm$ 7.1\%} & \multicolumn{1}{l}{51\%\;\;\, $\pm$ 13.7\%} & \multicolumn{1}{l}{54.9\% $\pm$ 13.7\%} & \multicolumn{1}{l}{\boldmath{}\textbf{90.2\% $\pm$ 8.2\%}\unboldmath{}} & \multicolumn{1}{l}{74.5\% $\pm$ 6.1\%} \\
          & Holdout & \multicolumn{1}{l}{100\%} & \multicolumn{1}{l}{60\%} & \multicolumn{1}{l}{100\%} & \multicolumn{1}{l}{100\%} & \multicolumn{1}{l}{0\%} & \multicolumn{1}{l}{70\%} & \multicolumn{1}{l}{100\%} & \multicolumn{1}{l}{60\%} & \multicolumn{1}{l}{0\%} & \multicolumn{1}{l}{63.6\%} & \multicolumn{1}{l}{100\%} & \multicolumn{1}{l}{100\%} & \multicolumn{1}{l}{50\%} & \multicolumn{1}{l}{65\%} & \multicolumn{1}{l}{\textbf{100\%}} & \multicolumn{1}{l}{80\%} \\
          {Mind Research} & Training & \multicolumn{1}{l}{100\%} & \multicolumn{1}{l}{N/A} & \multicolumn{1}{l}{88.2\% $\pm$ 11.5\%} & \multicolumn{1}{l}{100\%} & \multicolumn{1}{l}{0\%} & \multicolumn{1}{l}{N/A} & \multicolumn{1}{l}{84.6\% $\pm$ 12.9\%} & \multicolumn{1}{l}{23.1\% $\pm$ 15.1\%} & \multicolumn{1}{l}{0\%} & \multicolumn{1}{l}{N/A} & \multicolumn{1}{l}{84.6\% $\pm$ 12.9\%} & \multicolumn{1}{l}{100\%} & \multicolumn{1}{l}{56.7\% $\pm$ 17.7\%} & \multicolumn{1}{l}{N/A} & \multicolumn{1}{l}{\boldmath{}\textbf{86.7\% $\pm$ 12.2\%}\unboldmath{}} & \multicolumn{1}{l}{66.7\% $\pm$ 16.9\%} \\
      Network (MRN)   & Holdout & \multicolumn{1}{l}{100\%} & \multicolumn{1}{l}{N/A} & \multicolumn{1}{l}{67\%} & \multicolumn{1}{l}{100\%} & \multicolumn{1}{l}{0\%} & \multicolumn{1}{l}{N/A} & \multicolumn{1}{l}{71\%} & \multicolumn{1}{l}{14\%} & \multicolumn{1}{l}{0\%} & \multicolumn{1}{l}{N/A} & \multicolumn{1}{l}{63\%} & \multicolumn{1}{l}{100\%} & \multicolumn{1}{l}{56\%} & \multicolumn{1}{l}{N/A} & \multicolumn{1}{l}{\textbf{68.8\%}} & \multicolumn{1}{l}{62.5\%} \\
   \multirow{2}[0]{*}{Stanford} & Training & \multicolumn{1}{l}{100\%} & \multicolumn{1}{l}{73.3\% $\pm$ 15.8\%} & \multicolumn{1}{l}{80\%\;\,\, $\pm$ 14.3\%} & \multicolumn{1}{l}{100\%} & \multicolumn{1}{l}{0\%} & \multicolumn{1}{l}{66.7\% $\pm$ 16.9\%} & \multicolumn{1}{l}{86.7\% $\pm$ 12.2\%} & \multicolumn{1}{l}{33.3\% $\pm$ 16.9\%} & \multicolumn{1}{l}{0\%} & \multicolumn{1}{l}{71.4\% $\pm$ 16.2\%} & \multicolumn{1}{l}{81.3\% $\pm$ 14\%} & \multicolumn{1}{l}{100\%} & \multicolumn{1}{l}{50\%\;\;\, $\pm$ 17.9\%} & \multicolumn{1}{l}{70\% \;\;\,$\pm$ 16.4\%} & \multicolumn{1}{l}{\boldmath{}\textbf{83.3\% $\pm$ 13.3\%}\unboldmath{}} & \multicolumn{1}{l}{66.7\% $\pm$ 16.9\%} \\
          & Holdout & \multicolumn{1}{l}{100\%} & \multicolumn{1}{l}{80\%} & \multicolumn{1}{l}{80\%} & \multicolumn{1}{l}{100\%} & \multicolumn{1}{l}{0\%} & \multicolumn{1}{l}{40\%} & \multicolumn{1}{l}{60\%} & \multicolumn{1}{l}{0\%} & \multicolumn{1}{l}{0\%} & \multicolumn{1}{l}{67\%} & \multicolumn{1}{l}{75\%} & \multicolumn{1}{l}{0\%} & \multicolumn{1}{l}{50\%} & \multicolumn{1}{l}{60\%} & \multicolumn{1}{l}{\textbf{70\%}} & \multicolumn{1}{l}{50\%} \\
   University of & Training  & \multicolumn{1}{l}{100\%} & 69.2\% $\pm$ 20.2\% & \multicolumn{1}{l}{100\%} & \multicolumn{1}{l}{100\%} & \multicolumn{1}{l}{0\%} & \multicolumn{1}{l}{42.9\% $\pm$ 21.7\%} & \multicolumn{1}{l}{71.4\% $\pm$ 19.8\%} & \multicolumn{1}{l}{14.3\% $\pm$ 15.3\%} & \multicolumn{1}{l}{0\%} & 42.9\% $\pm$ 21.7\% & \multicolumn{1}{l}{100\%} & \multicolumn{1}{l}{100\%} & 65\% \;\;\,$\pm$ 20.9\% & 60\%\;\;\, $\pm$ 21.5\% & \boldmath{}\textbf{90\%\;\;\, $\pm$ 13.1\%}\unboldmath{} & 70\%\;\;\, $\pm$ 20.1\% \\
    Michigan (UM\_2) & Holdout & \multicolumn{1}{l}{100\%} & \multicolumn{1}{l}{66.6\%} & \multicolumn{1}{l}{100\%} & \multicolumn{1}{l}{83.3\%} & \multicolumn{1}{l}{0\%} & \multicolumn{1}{l}{25\%} & \multicolumn{1}{l}{50\%} & \multicolumn{1}{l}{0\%} & \multicolumn{1}{l}{0\%} & \multicolumn{1}{l}{20\%} & \multicolumn{1}{l}{100\%} & \multicolumn{1}{l}{0\%} & \multicolumn{1}{l}{60\%} & \multicolumn{1}{l}{30\%} & \multicolumn{1}{l}{\textbf{80\%}} & \multicolumn{1}{l}{50\%} \\
    \bottomrule
    \end{tabular}
    }

\end{table*}

\begin{table*}[h]
  
  \centering
    \caption{Dataset Summary}  \label{tbl:dsinfo}%
  \footnotesize
  \tabcolsep 1pt
  {\fontsize{6.75}{7}\selectfont
    \begin{tabular}{rrrrrrrrrrrcrrr}
    \toprule
    \multirow{2}[0]{*}{Dataset} & \multicolumn{1}{c}{Experimental} & \multirow{2}[0]{*}{Pulse Sequence} & \multicolumn{1}{c}{\multirow{2}[0]{*}{Patient Disorder}} & \multicolumn{1}{c}{\multirow{2}[0]{*}{Source}} & \multicolumn{3}{c}{\multirow{2}[-5]{*}{Training Data}} & \multicolumn{3}{c}{\multirow{2}[-5]{*}{Holdout Data}} & \multicolumn{1}{c}{\multirow{2}[-5]{*}{fMRI Dimensionality}} & \multicolumn{2}{c}{\multirow{2}[-5]{*}{Parameters}} & \multicolumn{1}{c}{\multirow{2}[-5]{*}{Age Range}} \\
          &\multicolumn{1}{c}{Design} &  &  &  & total & patients & controls & total & patients & controls & ($L \times W \times H \times T $) & LLE ($r$,$d$) & PCA ($d$) & [min,mean,max] \\
              \toprule
    Beijing (Peking\_3) & resting-state & EPI   & ADHD  & ADHD200 & 30    & 15    & 15    & 12    & 4     & 8     & $57\times 68 \times 42 \times 236$ & (2,236) & 236    & [11, 13.24, 16] \\
    COBRE & resting-state & EPI   & Schizophrenia & COBRE  & 51    & 25    & 26    & 20    & 10    & 10    & $53 \times 62 \times 52 \times 150$ & (2,150) & 150   & [18, 37.45, 65] \\
    Mind Research Network (MRN) & block-design & EPI   & Schizophrenia & MCIC & 30    & 13    & 17    & 16    & 7     & 9     & $57 \times 68 \times 38 \times 177$ & (2,177) & 177    & [18, 34.15, 60] \\
    Stanford & resting-state & Spiral-IN/OUT EPI  & Autism & ABIDE & 30    & 15    & 15    & 10    & 5     & 5     & $63 \times 75 \times 42 \times 180$ & (2,60) & 60   & [7.52, 9.95, 12.93] \\
    University of Michigan (UM\_2) &  resting-state & Spiral-IN EPI & Autism & ABIDE & 20 & 7 & 13 & 10 & 4 & 6 & $57 \times 68 \times 63 \times 300$ & (2,222) & 222 & [13.1, 16.26, 28.8] \\
    \bottomrule
    \end{tabular}
}
\end{table*}

\subsection[Feature \& Parameter Selection]{Feature \& Parameter Selection}\label{app:fs}
\textit{Hyper-parameter Selection} involves selecting the algorithm parameters that will be used to generate the reconstructed fMRI from LLE and PCA, respectively. Both PCA and LLE require one to specify the number of reconstruction dimensions $d$, where LLE's additional hyper-parameter, $K=(1+2r)^3-1$, selects the neighbouring voxels whose coordinates are on the boundary of a sphere with radius $r \in\mathbb{Z}_{+},r\geq1$ that is centred around the respective voxel.  For both PCA and LLE, the number of reconstruction dimensions is bounded by the number of time points. In the case of PCA, the optimal \textit{d} is often chosen by the proportion of variance captured by the first \textit{d} eigenvectors, expressed as $\frac{\sum_{i=1}^d\lambda_i}{\sum_{j=1}^T\lambda_j}$. For LLE, however, there is no analogous interpretation because the \textit{d} eigenvectors are uniformly spaced ``time points" on the {\it world line}~\cite{mink}. Thus cross-validation procedures are implemented (see following subsection for details) on the training data to systematically select the best hyper-parameters, iterating over $d \in\{1,\ldots,T\}$, where $d$ is generated on a log-scale from 1 to $T$.

 \paragraph[]{Sequential Forward Selection (SFS)\footnote{Implemented using MATLAB's (R2013a) \textit{sequentialfs} function}~\cite{sfs}} is a nonparametric method for measurement (feature) selection that starts with an empty ``candidate", or near-optimal, set of ``time points". The method first finds the ``time point", defined as the second moment~\cite{bills}, with the highest classification accuracy and adds the corresponding image volume to the set. The method then finds an \textit{additional} ``time point" \comments{\deleted{(or volume)}} that \textit{strictly} improves the classification accuracy in \textit{conjunction} with the volume(s) whose ``time points" are already in the ``candidate" set, and terminates when no such volume can be found. Thus, SFS produces a near-optimal set of ``time points" (volumes) that distinguish the subject groups with high classification accuracy, where this set is determined by adding ``time points" to the near optimal set in a one-by-one fashion. It follows that the near-optimal set of ``time points" represent points in time during the scan that enable discrimination of patients from controls with high classification accuracy. Note that SFS is only used on the training data (see Section~\ref{sect:evalcrit}).  

\subsection{Classifier}
A slight abuse of notation is introduced by redefining variable $\mathbf{z} \in \mathbb{R}^{Vc}$ as the one-dimensional representation of the \textit{c} diagnostic volumes from the respective subject's reconstructed fMRI $\mathbf{Z}$ produced in the previous step. Here, it is assumed there are \textit{c} diagnostic time points and that $\mathcal{Z}$ is a random variable containing the collection of cohorts' reconstructed fMRI.

\paragraph[]{Fisher's Linear Discriminant (LDA)\footnote{Implemented using MATLAB's (R2013a) \textit{classify} function with the `diaglinear' argument to estimate the positive diagonal covariance matrix}~\cite{fisher1,fisher2}} is a linear classification rule~\cite{htf} that assumes both the patient and control class densities (at each voxel location) can be represented as multivariate Gaussians in three-dimensional space, each with some intrinsic curvature~\cite{gauss2,gauss3}. Each class density is expressed as
\begin{equation*}
{f_k(\mathbf{z}) = \frac{1}{(2\pi)^{\frac{Vc}{2}}|\boldsymbol\Sigma_k|^{\frac{1}{2}}}}{\Large e^{-\frac{1}{2}(\mathbf{z}-\boldsymbol{\mu}_k)^\intercal\boldsymbol\Sigma_k^{-1}(\mathbf{z}-\boldsymbol{\mu}_k)}}
\end{equation*}
\noindent where it is assumed the $k$ classes have a common covariance matrix-- i.e., $\boldsymbol\Sigma_k = \boldsymbol\Sigma,\,\forall k$. This assumption allows the log ratio between the posterior distribution of each class to form a decision boundary that lies between patients (class 1) and controls (class 0), written as $P(Y=0|\mathcal{Z}=\mathbf{z}) = P(Y=1|\mathcal{Z}=\mathbf{z})$, which is linear in $\mathbf{z}$. LDA therefore calculates the $(Vc)$-dimensional hyperplane that best discriminates, or separates, the diagnostic volumes that have been determined using SFS with LOOCV on the respective dataset, for each class' reconstructed fMRI. It follows these diagnostic volumes contain spatial locations with sufficiently different patient and control class densities, where statistically-significantly different regions possess the requisite margin between the class densities such that they are perceptible~\cite{neuronuclear}.

\subsection{Evaluation Criteria}\label{sect:evalcrit}
\par Each dataset was split such that the proportion of patients in the training partition was near-equal to the proportion of controls, and the holdout dataset contained at least 10 cohorts. In all but two cases, the aforementioned criteria could not be met because the datasets were unbalanced. For these situations, the training dataset was constructed such that it was a representative sample of the overall data, with the remaining cohorts being assigned to the holdout dataset. The holdout set therefore contained group proportions that could deviate from the training set. The cohorts used to define the training and holdout partition for each dataset are provided in Table 3 of the Supplementary Information.

\par Performance was compared to the original fMRI, both before and after applying PCA for artefact reduction~\cite{nikos2}, and chance, which is defined as the proportion of the majority class on the respective dataset. The performance of the reconstruction methods are evaluated using a combination of Leave-One-Out Cross Validation (LOOCV) and holdout data classification performance~\cite{htf}.  LOOCV is used on the training data to find both the best reconstruction parameter $d$, and the diagnostic volumes that produce the highest accuracy for this $d$. The holdout data classification accuracies use the parameters found from performing LOOCV on the training data. Note: the cohorts in the holdout set are \textit{never} involved in determining the optimal hyper-parameter $d$, \textit{or} the diagnostic volumes produced from this $d$.

\begin{figure*}[htp!]
\centering
\begin{subfigure}[t]{7.25in}
\centering
\includegraphics{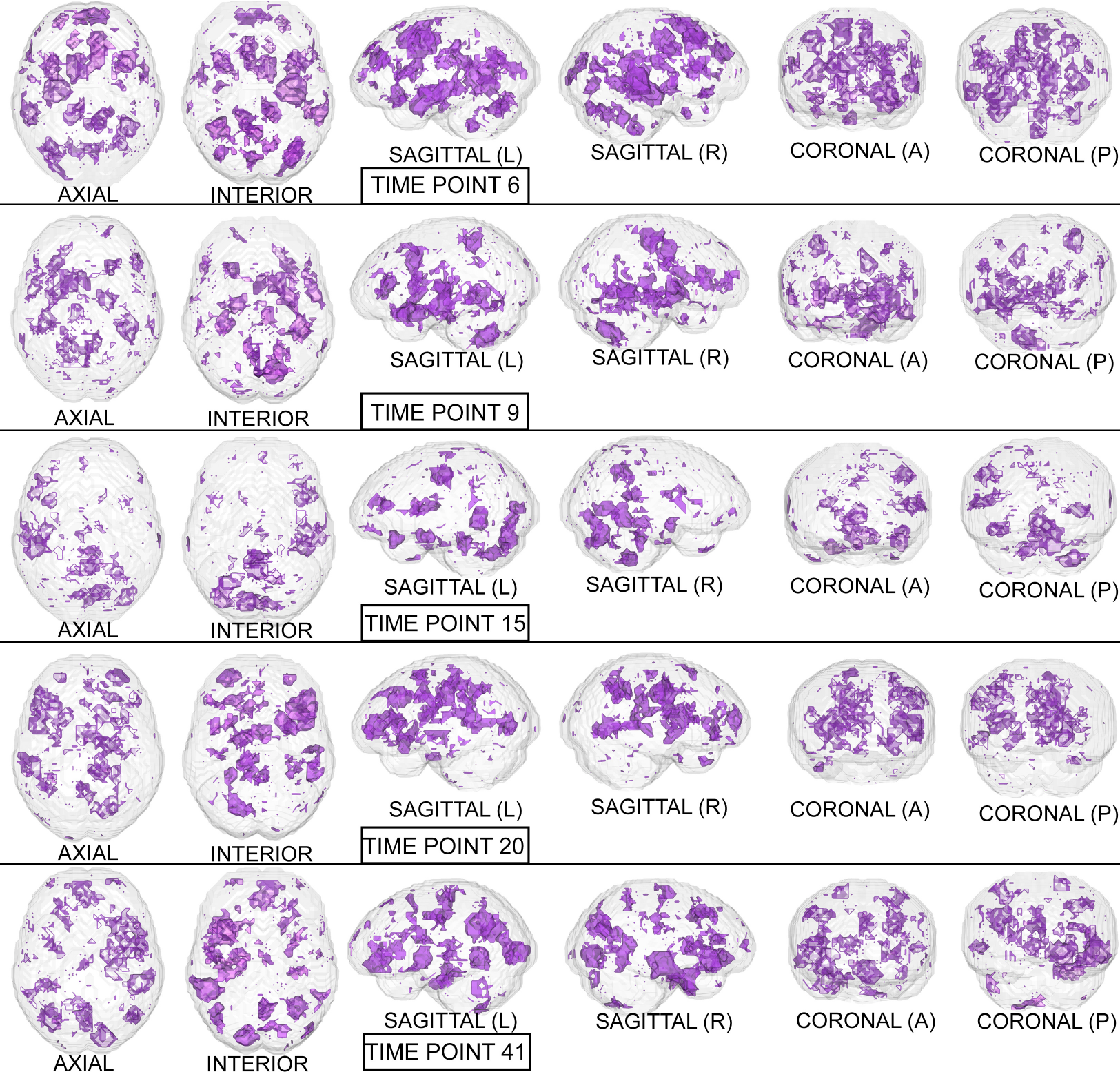}
\subcaption{Center of Biomedical Research Excellence (COBRE). Cohorts were at rest for the duration of the scan.}\label{subfig:cobrevol}
\end{subfigure}\\
\vfill \begin{subfigure}[t]{7.25in}
\centering
\includegraphics{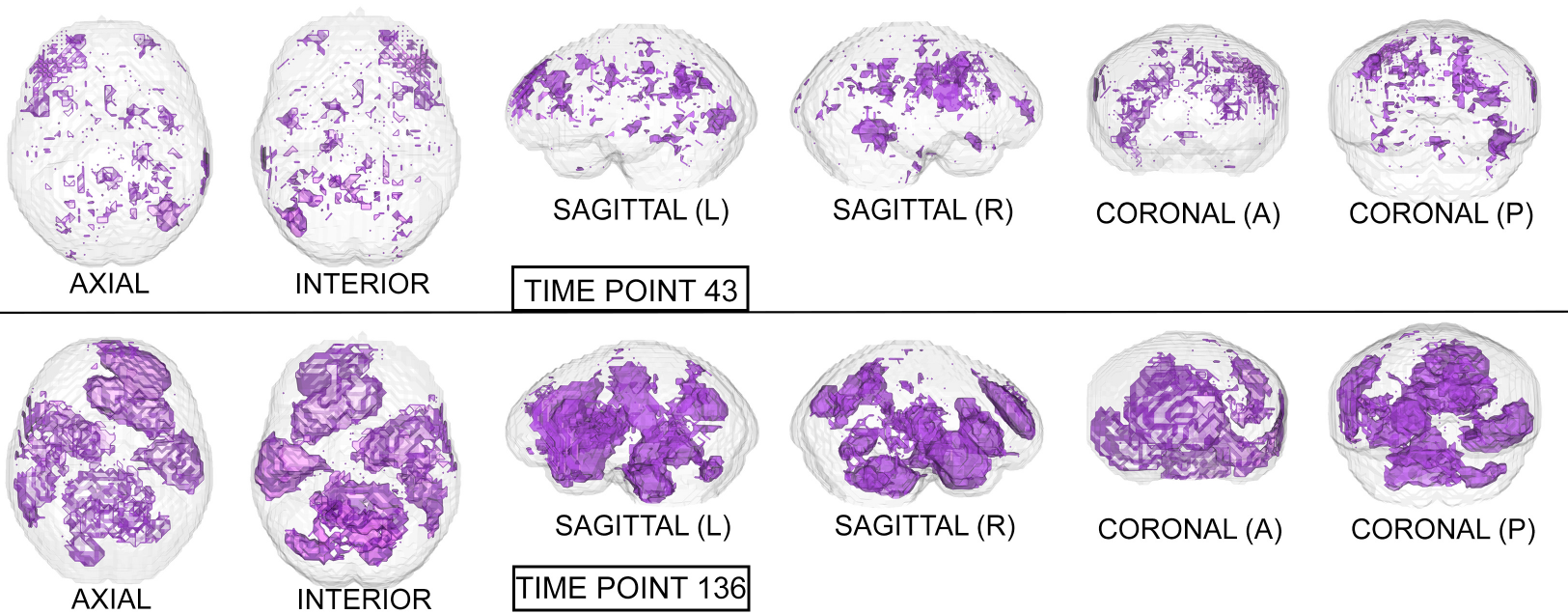}
\subcaption{Mind Research Network (MRN). Cohorts performed the Sternberg Item Recognition Paradigm (SIRP) task.}\label{subfig:mrnvol}
\end{subfigure}
\caption{\small Statistical maps illustrating the individual differences in mental activity (schizophrenic patients versus healthy controls) for the discriminative time points determined on the training partition. }\label{fig:schiz}
\end{figure*} 
\begin{figure*}[ht!]

\begin{subfigure}[t]{7.25in}
\centering
\includegraphics{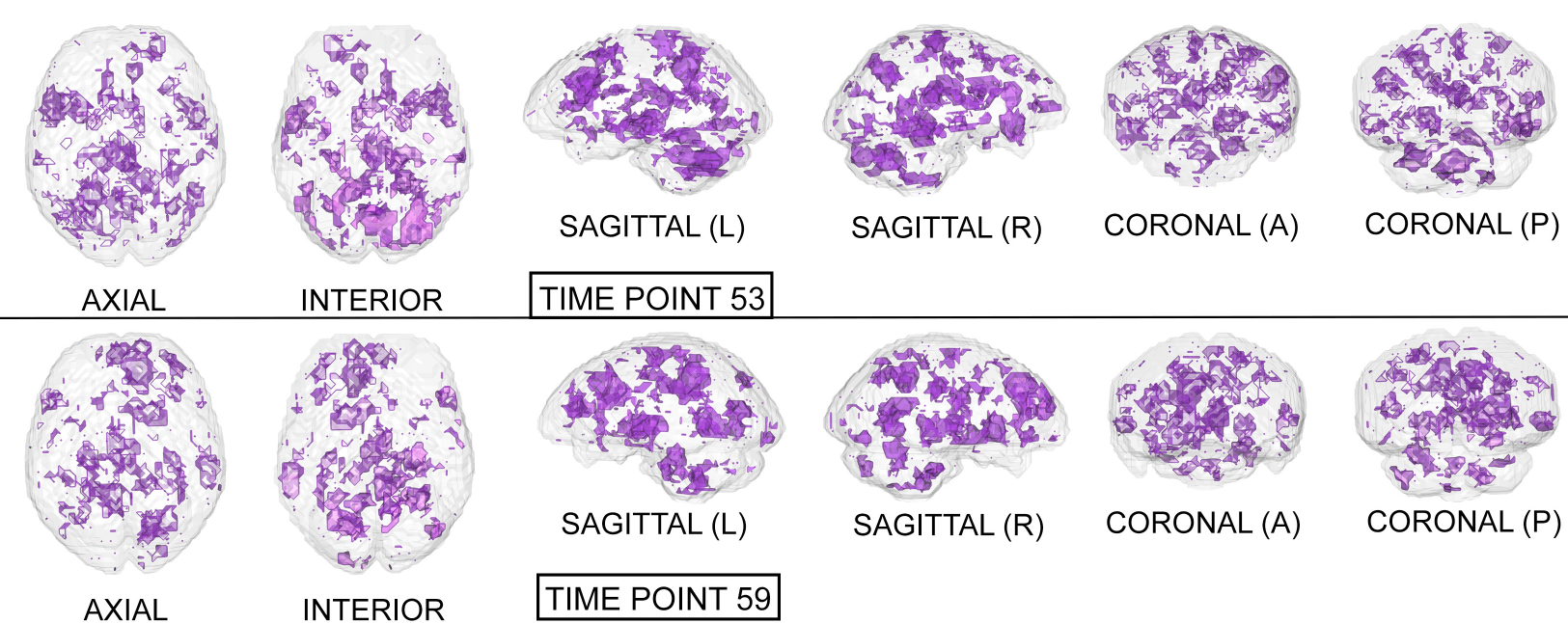}
\subcaption*{Beijing (Peking\_3) University.  Cohorts were at rest for the duration of the scan.}
\end{subfigure}
\caption{\small Statistical maps illustrating the individual differences in mental activity (ADHD patients versus healthy controls) for the discriminative time points determined on the training partition.}\label{fig:adhdvol}
\end{figure*}
\begin{figure*}[ht!]
\centering
\begin{subfigure}[t]{7.25in}
\centering
\includegraphics{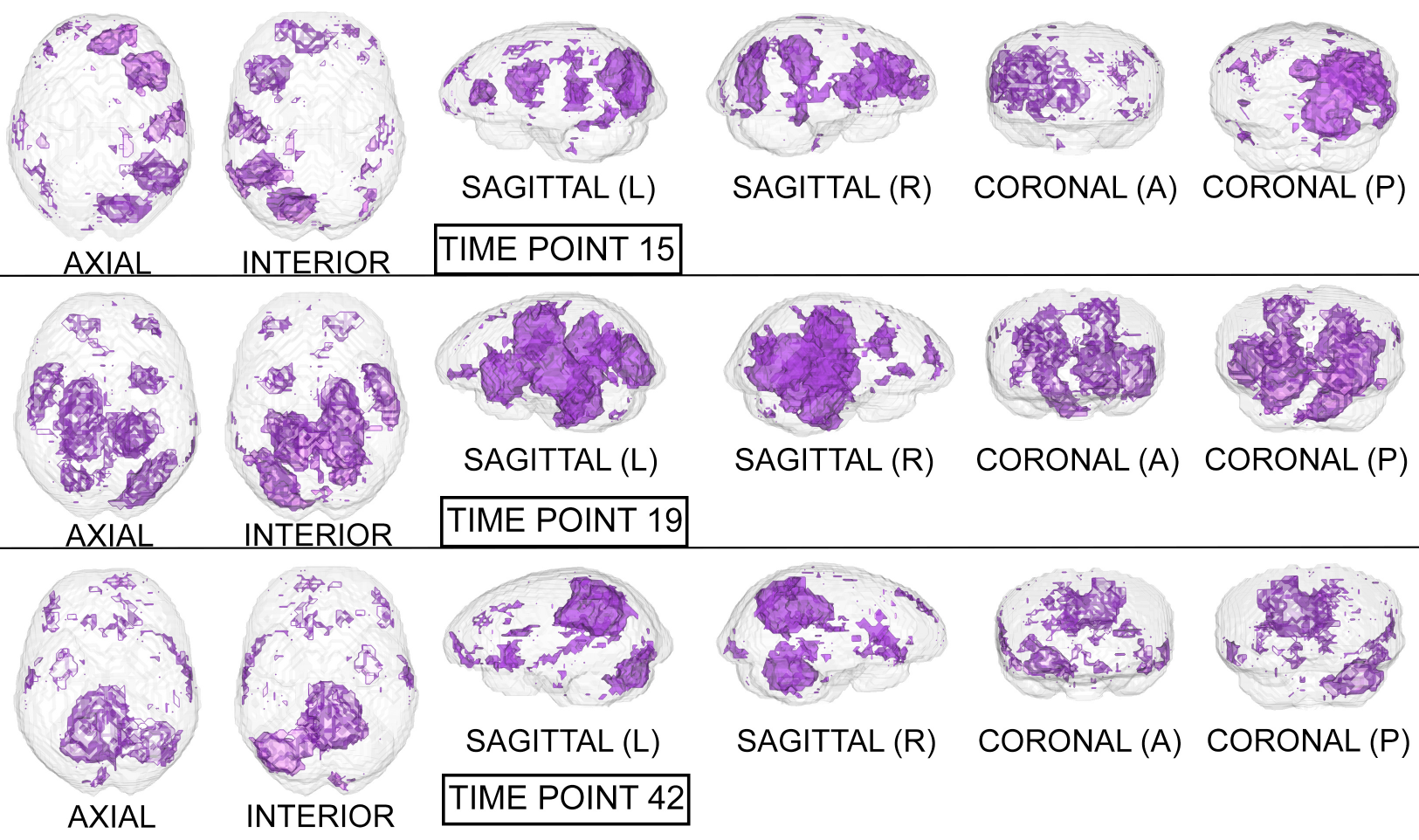}
\subcaption{Stanford University. Cohorts were at rest for the duration of the scan.}\label{subfig:stanvol}
\end{subfigure}
\begin{subfigure}[t]{7.25in}
\centering
\includegraphics{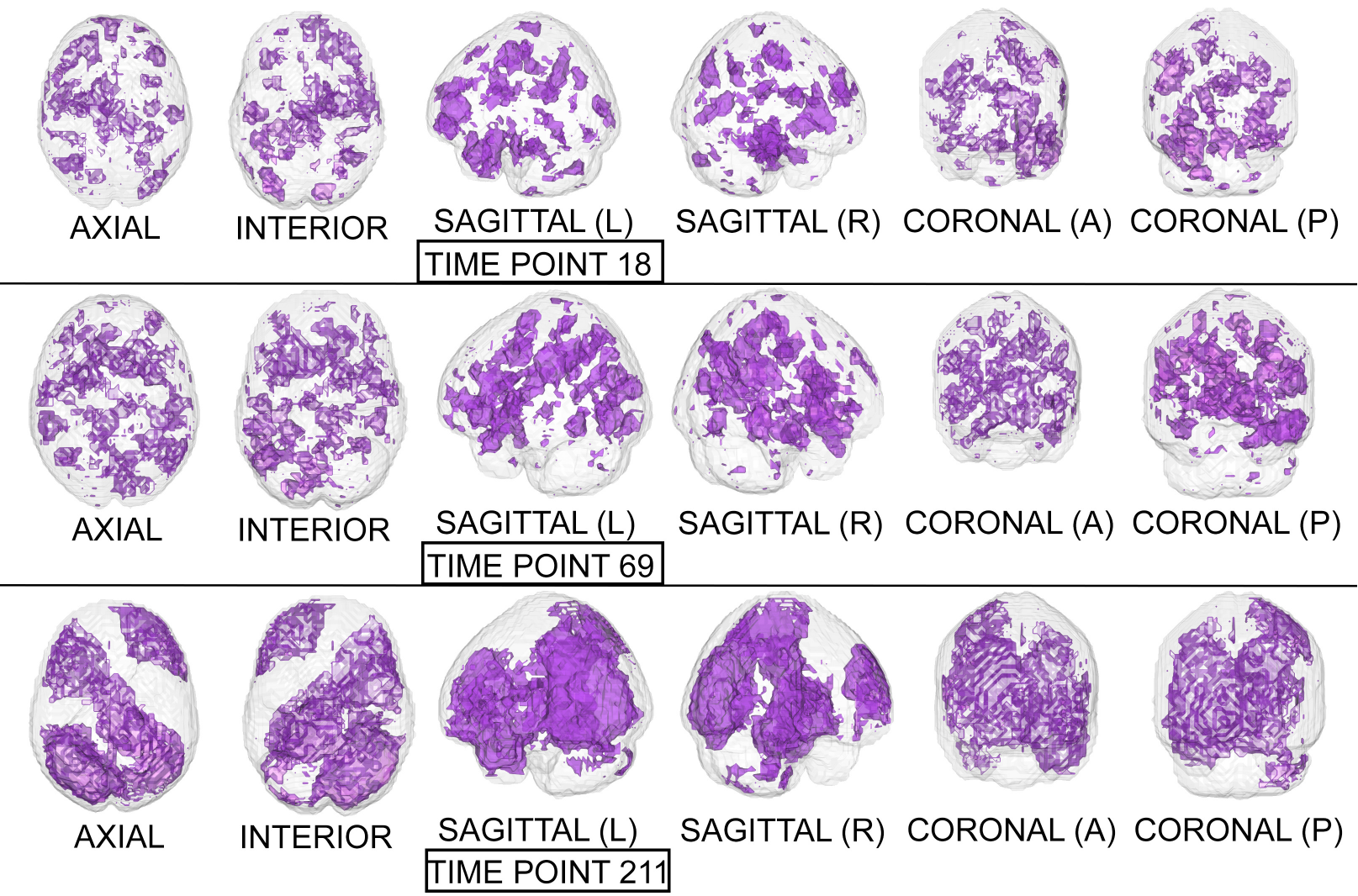}
\subcaption{University of Michigan (UM\_2). Cohorts were at rest for the duration of the scan.}\label{subfig:um2vol}
\end{subfigure}
\caption{\small The statistical maps illustrating the individual differences in mental activity (ASD patients versus healthy controls) for the discriminative time points determined on the training partition.}  \label{fig:asdvol}
\end{figure*}

\section{Results}
\subsection*{Performance \& Visualisation}
\par Tables~\ref{tbl:results} and \ref{tbl:dsinfo} demonstrate that Locally Linear Embedded fMRI can distinguish various mental disorders from healthy controls with high discriminatory power ($>$ 80\%); results for six additional resting-state datasets are provided in Table 2 of the Supplementary Information. The datasets contained healthy controls and patients with either Schizophrenia, Attention-Deficit Hyperactivity Disorder (ADHD), or Autism Spectrum Disorder (ASD). Ten of these datasets contained cohorts in the resting-state, with the remaining containing schizophrenic patients and healthy controls performing the Sternberg Item Recognition Paradigm (SIRP) task~\cite{sternberg}. For a description of the data sources and technical details, please consult the appendix.

 Given that performance on the training partition uses Leave One Out Cross Validation (LOOCV) to individually predict each subject's diagnosis, which is analogous to performing \textit{n} Bernoulli trials~\cite{bernoulli} (where \textit{n} is the number of cohorts), the training data's performance metrics can be interpreted as the mean of successes over \textit{n} binomially-distributed observations. To calculate the error of these estimates, we follow Laplace's approach of employing a normal distribution to estimate the error of binomially-distributed observations~\cite{laplace}. Given that discrimination performance on the holdout partition is determined in a one-time fashion, variance estimates are not applicable. 

\par The Harvard-Oxford Subcortical/Cortical and Cerebellum atlases~\cite{fsl} are used to identify the statistically-significant differences between patients and controls in the diagnostic volumes for the respective dataset, shown in Figures~\ref{fig:schiz},~\ref{fig:adhdvol}, and~\ref{fig:asdvol} (Figures 1, 2, 3, 4, 5, and 6 for datasets in the Supplementary Information). Each dataset's figure contains six different views of the statistically-significant differences at the time reflected by the respective time point(s); the coloured voxels at these time points denote statistically significant (\textit{p} < 0.05 uncorrected) physical differences between the patient and control groups, where these groups include cohorts from \textit{both} the training and holdout partitions. The proportion of significantly different voxels in each region, calculated by dividing the number of significantly different voxels by the total number of voxels in the respective region defined by the atlas, are provided in Tables 4 and 5 of the Supplementary Information.

\subsection*{Neurobiological Interpretation}

\paragraph{Schizophrenia}

\par David Ingvar \& G\"oran Franz\`en found that healthy controls exhibit increased flows in the prefrontal regions and decreased flow in the post central regions, and schizophrenic patients exhibited  the reversed pattern, with low flows prefrontally and high flows postcentrally~\cite{lancetschiz}. Furthermore, they noticed that a lower flow in the premotor and frontal regions was associated with symptoms of indifference, activity and autism, and a higher postcentral flow over the temporo-occipito-parietal regions was associated with disturbed cognition~\cite{lancetschiz}. Inspecting the statistical maps for the volumes in Figure~\ref{fig:schiz}, the significantly different areas seem to further substantiate Ingvar \& Franz\`en's observations, as there are significant differences in the various temporo-occipital-parietal regions in all of the volumes.

\par The Sternberg Item Recognition Paradigm (SIRP) task evaluates cohorts' short-term, or working, memory. Each of the seven tasks during the scan involves the subject memorising a set of objects, followed by presentation of a new object whose membership in this set is identified by a `yes` or `no`. The tasks therefore evaluate the hypothesised information processing differences between schizophrenics and healthy controls~\cite{classcalhoun}. It has been shown that the prefrontal and medial temporal regions are involved in encoding information, and it is believed the interactions between these regions are central to retrieval of stored information~\cite{memoryreview}. Figure~\ref{subfig:mrnvol}, especially volume 136, illustrates significant differences in the areas associated with the prefrontal and medial temporal regions; this is further supported by the fact that the accuracy on the holdout data rose from 68.8\% to 75\% when using only volume 136. Thus, it is possible that schizophrenia indeed affects the physical mechanisms associated with retrieving stored information, as these mechanisms are central to the SIRP task. The reconstruction method therefore successfully reveals physical differences associated with task performance between patients and controls, which are different from the resting-state differences for the same subject groups. 



\paragraph{Attention-Deficit Hyperactivity Disorder (ADHD)}

	
\par Similar to the goals of Ingvar \& Franz\`en~\cite{lancetschiz}, previous work used PET scans to compare the regional cerebral blood flow of children with Attention-Deficit Hyperactivity Disorder (ADHD) to healthy controls, where it was found that the disorder was associated with hypoperfusion in the striatal and posterior periventricular regions~\cite{lancetadhd}; these results provide biological evidence that is consistent with the canonical model for ADHD as a fronto-striatal deficient disorder. Figure~\ref{fig:adhdvol} shows significant differences in the various occipital, striatal, cerebellar and ventral regions of the brain. 

\begin{figure*}[htp!]
\centering
\begin{subfigure}[b]{2in}
\centering
\includegraphics[scale=0.85]{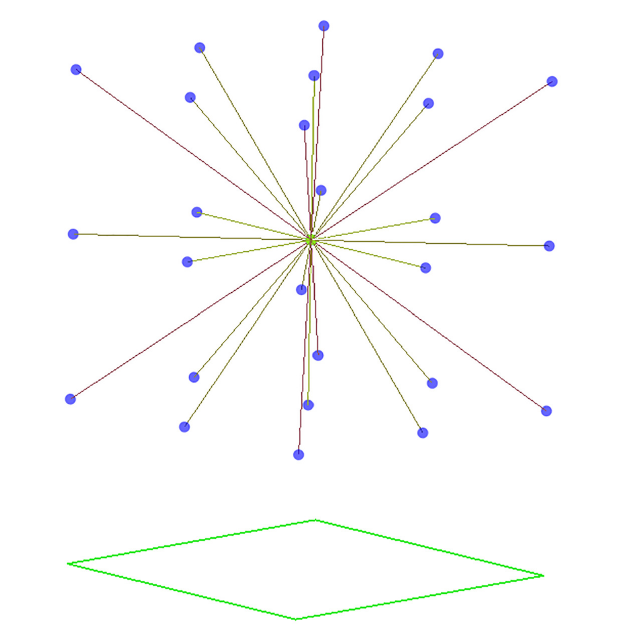}
\subcaption{\small An illustration~\cite{clrs} of the Cauchy stress tensor of a spherical body with radius $r=1$.}
\end{subfigure}\quad\quad\quad
\begin{subfigure}[b]{4.5in}
\centering
\includegraphics[scale=0.85]{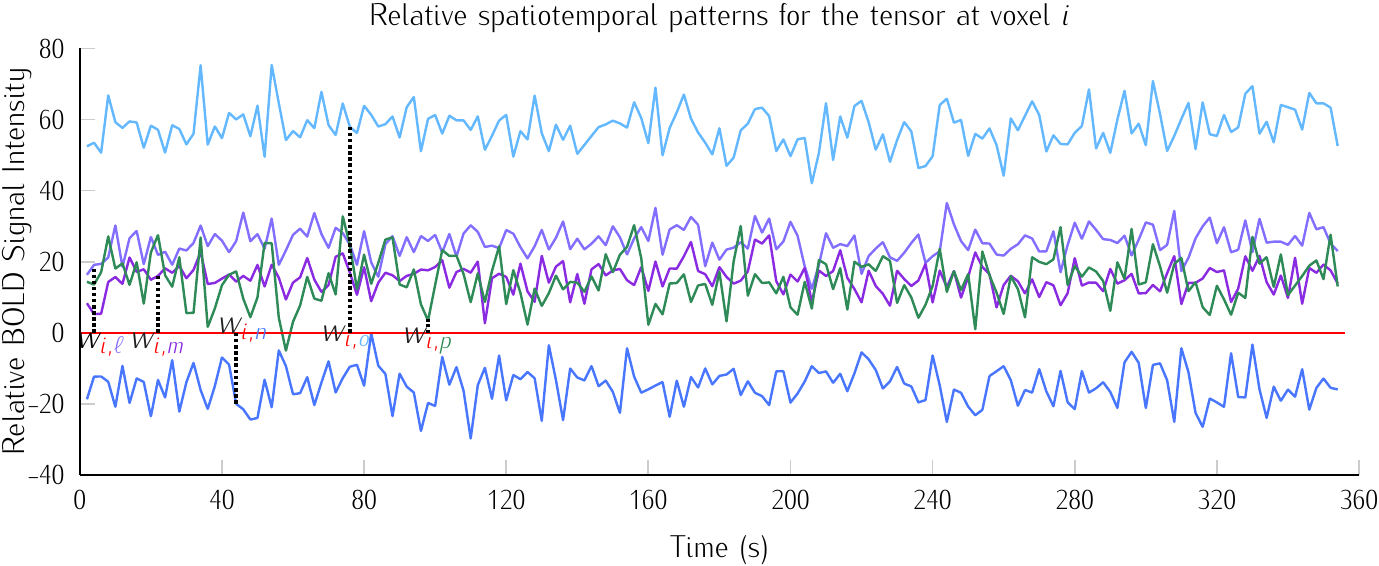}
\subcaption{\small The relative spatiotemporal patterns represent the surface forces relative to the \textit{deflexion axis}~\cite{dadmackay1,dadmackay2} at voxel $i$. The stress vectors (Lebesgue measures~\cite{lebesgue}) are the minimum-norm solution to the unitarily-constrained least-squares problem $\mathbf{G}_i\mathbf{w}_i=\mathbf{1}_{|\mathcal{N}(i)|}$, where $\mathbf{G}_i \in\mathbb{R}^{|\mathcal{N}(i)|\times|\mathcal{N}(i)|}$ is the local spatiotemporal covariance matrix (Lebesgue, $\ell^2$, or square-integrable~\cite{schmidt} space).}
\end{subfigure} 
\caption{The local geometry of the Cauchy stress tensor and its relative spatiotemporal patterns on topology~\cite{bourbaki} $\mathcal{X}$, which is defined on the Cartesian space with the Pythagorean distance metric.}
\label{fig:blackbody}
\end{figure*}
\paragraph{Autism Spectrum Disorder (ASD)}
	Similar to Ingvar \& Franz\`en's observations that schizophrenic patients and healthy controls had normal hemisphere flows~\cite{lancetschiz}, studies using PET to compare the regional cerebral blood flow of Autism Spectrum Disorder (ASD) patients to healthy controls observed normal metabolism and blood flow. Hypoperfusion in the temporal lobes, centred in the associative auditory and adjacent multimodal cortex~\cite{autismpsych}, was observed in autistic children. Furthermore, this temporal hypoperfusion was individually identifiable in 75\% of autistic children~\cite{autismpsych}. Figure~\ref{fig:asdvol} illustrates statistical differences in many areas of the temporal lobe.   

In summary, Locally Linear Embedding appears to have conserved spatiotemporal patterns in resting-state fMRI (and task-related responses) that are consistent with literature on regionally-specific abnormalities of cerebral activity in the psychiatric conditions used to assess classification performance.

\section{Discussion}

\par The multidisciplinary nature of this work undoubtedly introduces difficulty when discussing its motivations, which is the deployment of this methodology in a clinical setting. Such a goal imposes some conditions. First, while the results support deployment, the methodology must be further evaluated by trained clinicians well-versed in the etymology of the disease under investigation. More importantly, however, experimental designs are compulsory when discovering biological markers for disease; that is, patients must be subject to the same stimulus presentation at the same time during the scan in order to homogenise comparisons. With respect to resting-state fMRI, it is felt that a consensus is required to glean neurobiological insight that can generalise, which makes further discussion more appropriate for future work.

\par This exposition focused on Locally Linear Embedding as a promising and effective form of dimensionality reduction as a pre-processing step for the analysis of fMRI time-series. The approaches and aims of this form of pre-processing share a close relationship with other approaches in imaging neuroscience. These approaches include Independent Component Analysis, the use of Support Vector Machines (and regression) to classification problems (and prediction), and algorithms based on adaptive smoothing. In future work, it will be interesting to explore the formal connections between other approaches and assess their relative sensitivity in the context of the classification problems considered above.

One hundred and fourty-eight years after Darwin ascertained that mental activity invokes physical mechanisms in the brain~\cite{darwin1871descent}, the brain of man and ant alike are among the marvellous collection of atoms in the world. 

{\setcounter{secnumdepth}{-1}
\section{Acknowledgements}

\par This work is dedicated to the late Sam Roweis (1972-2010), Donald MacCrimmon MacKay (1922-1987) and his son David John Cameron MacKay (1967-2016). Additionally, many thanks to colleagues Ruitong Huang \& Dr. Csaba Szepesv\`ari, Joshua T. Vogelstein, Dr. Vincent D. Calhoun, Dr. Neil Lawrence, Dr. Bert Vogelstein, Dr. Michael Milham, Dr. Karl J. Friston, Dr. Klaas Enno Stephan, Dr. Nikos K. Logothetis, Dr. Christof Koch, and Dr. Yifan Hu of AT\&T Labs' Information Visualisation department for his assistance with GraphViz, which was used to depict the Cauchy Stress Tensor in three dimensions. 
}

\appendix
\setcounter{secnumdepth}{1}


{

\subsection{Locally Linear Embedding}\label{app:LLE}
\par fMRI contain \textit{T} uniformly-spaced time points, where time point $t \in \{1,\ldots,T\}$ describes a three-dimensional space comprised of $V = L \times W \times H$ voxels (volumetric elements); this is the \textit{global} description.  Each voxel is a volumetric measurement of the brain's physical mechanisms in both space and time \begin{wrapfigure}[9]{tr}{0.255\textwidth}
\includegraphics{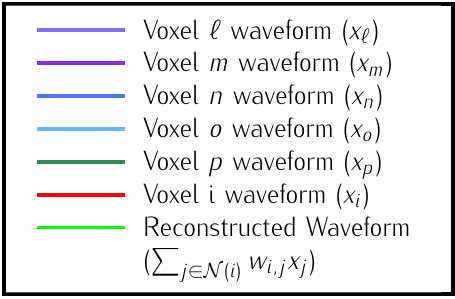}
\end{wrapfigure} -- i.e., every voxel represents a measurement at spatial coordinates  $(x,y,z)$ at time $t$, which can also be expressed as a four-tuple $(x,y,z,t)$. The BOLD contrast over \textit{T} uniformly-spaced time points can be viewed as a random variable, or measurable function,~\cite{bills} that is caused by local neuronal fluctuations. The terms global and local are only valid under a rigorous definition of three-dimensional space. A topological space requires specification of the relationships between points in a set and their respective open sets, which are defined as the sets that generalise the concept of an open interval in the real line while also providing a rigorous definition for nearness of points in this space~\cite{bourbaki}.\comments{ Define $\mathcal{X} = \big\{(x,y,z): x\in\{1,\ldots,L\}, y\in\{1,\ldots,W\}, z\in\{1,\ldots,H\}\big\}$ as the set containing the spatial coordinates of all voxels. Every point $(x_i,y_i,z_i) \in \mathcal{X}$ for $i \in \{1,\ldots,V\}$ is prescribed by the open set $\mathcal{N}(i) = \big\{(x,y,z): \sqrt{(x-x_i)^2+(y-y_i)^2+(z-z_i)^2} < r\big\}$, where $r$ defines the spatial radius. Thus, the open set of each spatial point generalises the open interval on the real line in Cartesian space. It follows that $\mathcal{X}$ is topological space that objectively describes physical reality in a homogenous, isotropic, and time-independent manner.  }

\begin{figure*}[t]
\begin{subfigure}[b]{3.5in}
\subcaption*{\;\;\;\;\;\;\;\;\;\;\;\;\;\;\;\;\;Original fMRI}
\centering
\includegraphics[scale=0.9]{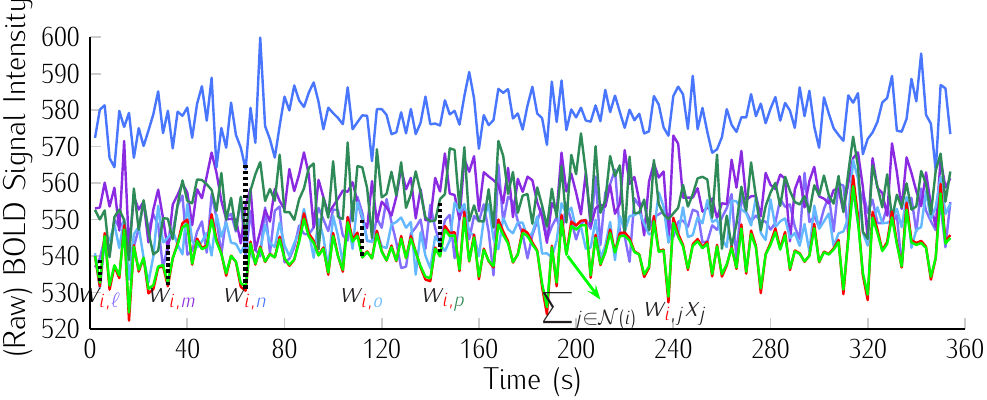}
\end{subfigure}\quad
\begin{subfigure}[b]{3.5in}
\subcaption*{\;\;\;\;\;\;\;\;\;\;\;\;Reconstructed fMRI\vspace{0.775em}}
\centering
\includegraphics[scale=0.9]{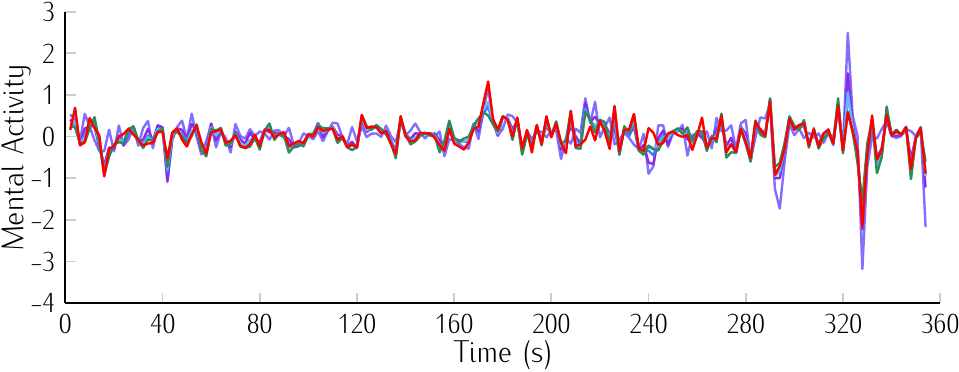}
\end{subfigure}
\caption{\small Illustrating the reconstruction of the waveform at voxel \textit{i}, located in the left cerebral cortex of a schizophrenic patient performing the Sternberg Item Recognition Paradigm (SIRP), as a linear combination of the Lebesgue measures~\cite{lebesgue} $w_{i,1},\ldots,w_{i,|{\cal N}(i)|}$ (defined on the Cauchy stress tensor) and the spatial patterns of the voxels on the boundary of the sphere, where spatial distance is defined using the Pythagorean distance metric; in this example, $r=2$. For comprehension, only the five nearest voxel waveforms are shown.}
\label{fig:llewvfrms}
\end{figure*}

\par Locally Linear Embedding (LLE)~\cite{lle1} is a congruent transformation~\cite{coxeter} that first extracts the local mathematical (geometric) structure of the data, and then performs a global optimisation that best conserves this local \textit{latent structure}~\cite{bacon}, or \textit{extension}~\cite{locke1690essay}. The application of LLE is challenging, primarily due to the implicit assumption that each data point and its neighbours lie on, or close to, a locally linear subspace~\cite{LLE2}; earlier applications of LLE to fMRI discarded spatial properties of the data in the neighbourhood selection step~\cite{mann}, thereby failing to preserve the inherent spatial configuration. In contrast, specifying the topological space $\mathcal{X}$ allows LLE to treat each data point's open set as its neighbourhood, thereby forgoing the neighbourhood selection step entirely. It follows that the benefit of LLE depends on the suitability of the distance metric used in determining the neighbourhood of each data point. Figures~\ref{fig:blackbody} and~\ref{fig:llewvfrms} suggest that the physical reality represented by topological space $\mathcal{X}$ is a collection of locally-linear subspaces whose properties are subject to change at any time. 


 \par LLE uses the \textit{local} description to construct the Cauchy stress tensor~\cite{cauchy} at every voxel, which proceeds as follows: Initially, the spatial pattern of each voxel \textit{i}, given by $\mathbf{x}_i\in\mathbb{R}^T$, is subtracted from both itself and the spatial patterns of its neighbours $\mathcal{N}(i)$, thereby allowing its ``zero waveform" (represented as $\mathbf{0}\in\mathbb{R}^T$) to serve as the imaginary plane, or \textit{deflexion axis}~\cite{dadmackay1,dadmackay2}, that divides the spherical body; the subtracted spatial patterns, given by (vector space) $\mathbf{C}_i = [ \mathbf{x}_j-\mathbf{x}_i,\ldots,\mathbf{x}_{j+|\mathcal{N}(i)|}-\mathbf{x}_i]  \in \mathbb{R}^{T \times |\mathcal{N}(i)|}$, represent the distances (over time) from the respective voxel's spatial location.

\par We then use the inner-product $\langle{(\mathbf{x}_j-\mathbf{x}_i),(\mathbf{x}_j-\mathbf{x}_i)}\rangle\in\mathbb{R}$ for $j \in \mathcal{N}(i)$ to compute the squared distances, which results in the real-symmetric positive-definite spatiotemporal covariance matrix $\mathbf{G}_i = \mathbf{C}_i^\intercal\mathbf{C}_i \in\mathbb{R}^{|\mathcal{N}(i)|\times |\mathcal{N}(i)|}$ for every voxel \textit{i}. Since the contact forces are inversely proportional to the squared differences represented in elements of $\mathbf{G}_i$~\cite{newton}, we use the Moore-Penrose inverse~\cite{pinv} to solve for the minimum-norm solution ($\mathbf{G}_i^+\mathbf{1}_{|\mathcal{N}(i)|} = \mathbf{w}_i$) to the constrained least squares problem $\mathbf{G}_i\mathbf{w}_i= \mathbf{1}_{|\mathcal{N}(i)|}$. Here, $\mathbf{1}_{|\mathcal{N}(i)|}\in\mathbb{R}^{|\mathcal{N}(i)|}$ is the unit-length direction vector in each of the $|\mathcal{N}(i)|$ directions, and weights $\mathbf{w}_i \in\mathbb{R}^{|\mathcal{N}(i)|}$ represent the stress vectors originating from voxel \textit{i}. By the spectral theorem, there exists an orthogonal basis $\mathbf{Q}_i \in\mathbb{R}^{d\times |\mathcal{N}(i)|}$ such that $\mathbf{G}_i = \mathbf{Q}_i^\intercal \mathbf{A} \mathbf{Q}_i$ and $\mathbf{Q}_i^\intercal\mathbf{Q}_i = \mathbf{I}\in\mathbb{R}^{|\mathcal{N}(i)|\times |\mathcal{N}(i)|}$~\cite{sylvester}; thus the stress vectors represented by $\mathbf{w}_i$ are shift, rotation, and translation invariant in the space defined on basis $\mathbf{Q}_i$. The substantial overlap between spatially-adjacent voxels' tensors precludes independently calculating orthogonal bases $\mathbf{Q}_1,\ldots,\mathbf{Q}_V$, as this defines each tensor's basis on separate vector spaces. \comments{We therefore seek a way to retain the spatially-invariant properties of \textit{all} voxels' Cauchy stress tensors by finding a \textit{global} coordinate system that treats space as a measurable function~\cite{bills} of time, where all voxels' measurements taken at time \textit{t} lie on a \textit{V}-dimensional basis vector, which is also one degree of freedom in space and time.} 

\par The stress vector $\mathbf{w}_i$ for each voxel's Cauchy stress tensor, $\mathbf{Q}_i$, represents the spatially-invariant properties of the three-dimensional forces applied over the duration of the scan in the respective subspace. LLE retains these spatially-local invariant properties by constructing the (global) adjacency matrix $\mathbf{W} \in\mathbb{R}^{V\times V}$ such that $\mathbf{W}_{i,j} \neq 0 \iff j \in \mathcal{N}(i) \iff \mathbf{W}_{j,i} \neq 0 \iff i \in \mathcal{N}(j)$; by construction $\mathbf{W}$ contains every voxel's stress vectors, which were determined using the Cauchy stress tensor. Given that the stress vectors' invariant properties are determined using squared distances, LLE then computes the squared distances between every voxel in the (global) normalised Laplacian matrix~\cite{chung} $(\mathbf{I} - \mathbf{W}) \in \mathbb{R}^{V\times V}$. This is expressed by $\mathbf{M} = (\mathbf{I} - \mathbf{W})^\intercal(\mathbf{I} - \mathbf{W})\mathbf{1}_{V}$, where $(\mathbf{I}-\mathbf{W})\mathbf{1}_V=\mathbf{0}\in\mathbb{R}^{V}$~\cite{lawrence}. By construction, $\mathbf{M}$ is symmetric and positive-definite, which means there exists an orthogonal basis $\mathbf{Z} \in \mathbb{R}^{V \times (d+1)}$ such that  $(\mathbf{I} - \mathbf{W})^\intercal(\mathbf{I} - \mathbf{W})\mathbf{1}_{V} \equiv \mathbf{Z}\Lambda\mathbf{Z}^\intercal$~\cite{drmackay}, where $\mathbf{1}_V$ represents the \textit{global} unit vector in three-dimensional space.\footnote{$\mathbf{1}_V$ is also the $d+1$ eigenvector of $\mathbf{Z}$, which is discarded after the global optimisation.} Since the Cauchy stress tensor is of second order, LLE uses Rayleigh's variational principle~\cite{rayleigh,courant} to calculate the resonance frequencies that best preserve the geometry of the \textit{deflexion axis} at every voxel's stress tensor. These are given by the bottom (d+1) eigenvectors, each of which represent one degree of freedom in space and time. LLE therefore uses a \textit{global} optimisation to embed the relative measurements of every voxel's Cauchy stress tensor~\cite{cauchy} in a global coordinate system $\mathbf{Z}^{V \times (d+1)}$ that conserves quantities over \textit{both} space and time~\cite{friston}, where this coordinate system contains a mechanical system~\cite{drkoch} in static equilibrium~\cite{caratheodory}. Applying LLE to fMRI data can therefore be viewed as using Carl Friedrich Gau{\ss}' Principle of Least Constraint~\cite{gauss} to determine the true motion of the mechanical system defined on the topology $\mathcal{X}$, where the Cauchy stress tensor~\cite{cauchy} allows preservation of the intrinsic local Gaussian curvature~\cite{gauss2,gauss3} in space and time.

\ifCLASSOPTIONcaptionsoff
  \newpage
\fi


\clearpage
\begin{thebibliography}{1}

\bibitem{pcafmri}
{\sc Andersen, A.~H., Gash, D.~M., and Avison, M.~J.}
\newblock Principal Component Analysis of the dynamic response measured by
  fMRI: a generalized linear systems framework.
\newblock {\em Magnetic Resonance Imaging}, 17(6):795 -- 815, 1999.

\bibitem{archibald1914}
{\sc Archibald, R.~C.}
\newblock Time as a fourth dimension.
\newblock {\em Bulletin of the American Mathematical Society 20}, (8):409--412, 1914.


\bibitem{bacon}
{\sc Bacon, F.}
\newblock {\em Francisci de Verulamio, summi Angliae cancellarii, Instauratio
  magna}.
\newblock apud Joannem Billium, Typographum Regium, 1620.



\bibitem{bernoulli}
{\sc Bernoulli, J.}
\newblock {\em Ars Conjectandi}.
\newblock Impensis Thurnisiorum, Fraetum, 1713.

\bibitem{bishop}
{\sc Bishop, C.~M.}
\newblock{\em Pattern Recognition and Machine Learning}.
\newblock{Springer, 2006}.

\bibitem{bills}
{\sc Billingsley, P.}
\newblock{\em Probability and measure}.
\newblock{Wiley, 1979}.

\bibitem{bourbaki}
{\sc Bourbaki, N.}
\newblock{\em {\'{E}l\'ements de math\'ematique. {T}opologie alg\'ebrique. {C}hapitres 1 \`a 4}}.
\newblock Springer-Verlag Berlin Heidelberg, Paris, 2016.
 
\bibitem{caratheodory}
{\sc Carath{\'e}odory, C.}
\newblock Untersuchungen {\"u}ber die grundlagen der thermodynamik.
\newblock {\em Mathematische Annalen}, 67(3):355--386, 1909.

\bibitem{cauchy}
{\sc Cauchy, A.~L.}
\newblock De la pression ou tension dans un corps solide.
\newblock {\em Exercices de Math\`ematiques} 2 (1827), 42--56.

\bibitem{chung}
{\sc Chung, F.~R.~K.}
\newblock {\em Spectral Graph Theory}.
\newblock American Mathematical Society, 1997.

\bibitem{clrs}
{\sc Cormen, T.~H., Leiserson, C.~E., Rivest, R.~L., and Stein, C.}
\newblock {\em Introduction to Algorithms (2. ed.)}.
\newblock MIT Press, 2001.

\bibitem{courant}
{\sc Courant, R.}
\newblock Variational methods for the solution of problems of equilibrium and vibrations.
\newblock {\em Bulletin of the American Mathematical Society 49}, (1):1--23, 1943.

\bibitem{coxeter}
{\sc Coxeter, H.~S.~M.}
\newblock {\em Introduction to geometry (2. ed.)}.
\newblock{Wiley, 1989}.

\bibitem{darwin1871descent}
{\sc Darwin, C.}
\newblock {\em The Descent of Man}.
\newblock London: Murray, 1871.

\bibitem{dadmackay2}
{\sc Deeley, E.~M. and MacKay, D.~M.}
\newblock Multiplication and division by electronic-analogue methods
\newblock {\em Nature 163}, 4147 (4 1949), 650--650

\bibitem{classcalhoun}
{\sc Demirci, O., Clark, V.~P., Magnotta, V.~A., Andreasen, N.~C., Lauriello,
  J., Kiehl, K.~A., Pearlson, G.~D., and Calhoun, V.~D.}
\newblock A review of challenges in the use of fMRI for disease classification.
\newblock {\em Brain Imaging Behav 2}, 3 (Sep 2008), 147--226.

\bibitem{fisher1}
{\sc Fisher, R.~A.}
\newblock {The Use of Multiple Measurements in Taxonomic Problems}.
\newblock {\em Annals of Eugenics 7}, 2 (1936), 179--188.

\bibitem{fisher2}
{\sc Fisher, R.~A.}
\newblock {The Statistical Utilization of Multiple Measurements}.
\newblock {\em Annals of Eugenics 8}, 4 (1938), 376--386.


\bibitem{friston}
{\sc Friston, K.~J.}
\newblock {The free-energy principle: a unified brain theory?}
\newblock {\em Nature Reviews Neuroscience 11}, 2 (2010), 127--138.

\bibitem{compkern}
{\sc Garabedian, P.~R., Schiffer, M.}
\newblock On existence theorems of potential theory and conformal mapping.
\newblock {\em Annals of Mathematics}, 52(1):164--187, 1950.

\bibitem{gauss2}
{\sc Gau{\ss}, C.~F.}
\newblock {\em Disquisitiones generales circa superficies curvas}.
\newblock Typis Ditericianis, Germany, 1828.

\bibitem{gauss3}
{\sc Gau{\ss}, C.~F.}  
\newblock {\em Theoria motus corporum coelestium in sectionibus conicis solem ambientium}.
\newblock Perthes et Besser, 1809.


\bibitem{gauss}
{\sc Gau{\ss}, C.~F.}
\newblock Ueber ein allgemeines grundgesetz der mechanik.
\newblock {\em Journal f\"ur die reine und angewandte Mathematik\/} (1829),
  232--235.
  
\bibitem{vsepr}
{\sc Gillespie, R.~J.}
\newblock Fifty years of the vsepr model.
\newblock {\em Coordination Chemistry Reviews}, 252(12--14):1315--1327, 7 2008.

\bibitem{gregory}
{\sc Gregory, J.}
\newblock {\em Geometriae pars universalis.}
\newblock  Patavii: Typis heredum Pauli Frambotti, Italy, 1668.

\bibitem{htf}
{\sc Hastie, T., Tibshirani, R., and Friedman, J.}
\newblock {\em {The Elements of Statistical Learning: Data Mining, Inference,
  and Prediction, Second Edition}}, 2nd ed. 2009. corr. 3rd printing 5th
  printing.~ed.
\newblock Springer Series in Statistics. Springer, Sept. 2009.

\bibitem{hilbert}
{\sc Hilbert,~D.}
\newblock {{Ueber die stetige Abbildung einer Line auf ein Fl\"{a}chenst\"{u}ck}}
\newblock {\em Mathematische Annalen 3} 38 (1891), 459--460

\bibitem{lancetschiz}
{\sc Ingvar, D.~H., and Franz{\'e}n, G.}
\newblock Distribution of cerebral activity in chronic schizophrenia.
\newblock {\em The Lancet 304}, 7895 (12 1974), 1484--1486.

\bibitem{lancetlassingvar}
{\sc Ingvar, D.~H., and Lassen, N.~A.}
\newblock Quantitative determination of regional cerebral blood-flow in man.
\newblock {\em The Lancet 278}, 7206 (10 1961), 806--807.

\bibitem{fsl}
{\sc Jenkinson, M., Beckmann, C.~F., Behrens, T. E.~J., Woolrich, M.~W., and
  Smith, S.~M.}
\newblock Fsl.
\newblock {\em Neuroimage 62}, 2 (Aug 2012), 782--790.

\bibitem{pcabook}
{\sc {Jolliffe}, I.}
\newblock {\em Principal component analysis}.
\newblock Springer Verlag, New York, 2002.



\bibitem{drkoch}
{\sc {Koch}, C., and Hepp, K.}
 \newblock{Quantum mechanics in the brain}
\newblock{\em Nature 440}, 7084 (2006), 611-612.

\bibitem{laplace}
{\sc Laplace, P.}
\newblock {\em Th{\'e}orie analytique des probabilit{\'e}s}.
\newblock Courcier, Paris, 1812.

\bibitem{lasseningvar}
{\sc Lassen, N.~A., and Ingvar, D.~H.}; {\sc Potchen, E.~J.,  McCready, V.~R.} (Eds.)
\newblock Radioisotopic assessment of regional cerebral blood flow.
\newblock {\em Progress in Nuclear Medicine 1: Neuro-Nuclear Medicine}.
\newblock University Park Press, 1972. 376--409


\bibitem{lawrence}
{\sc Lawrence, N.~D.}
\newblock A unifying probabilistic perspective for spectral dimensionality
  reduction: Insights and new models.
\newblock {\em Journal of Machine Learning Research 13\/} (June 2012), 1609--1638.

\bibitem{lebesgue}
{\sc Lebesgue, H.}
\newblock {\em Le{\c{c}}ons sur l'int{\'e}gration et la recherche des fonctions
  primitives: profess{\'e}es au Coll{\`e}ge de France}.
\newblock Collection de monographies sur la th{\'e}orie des fonctions.
  Gauthier-Villars, 1904.

\bibitem{liddle}
{\sc Liddle, P.~F., Friston, K.~J., Frith, C.~D., Hirsch, S.~R., Jones, T., and  Frackowiak, R.~S.}
\newblock Patterns of cerebral blood flow in schizophrenia.
\newblock {\em The British Journal of Psychiatry 160}, 2 (1992), 179--86.

\bibitem{locke1690essay}
{\sc Locke, J.}
\newblock {\em An Essay Concerning Humane Understanding}.
\newblock Thomas Basset, 1690.

\bibitem{nikos2}
{\sc Logothetis, N., Pauls, J., Augath, M., Trinath, T., and Oeltermann, A..}
\newblock Neurophysiological investigation of the basis of the fMRI signal.
\newblock {\em Nature 412}, 6843 (7 2001), 150--157.

\bibitem{natmrirev}
{\sc Logothetis, N.~K.}
\newblock What we can do and what we cannot do with fMRI.
\newblock {\em Nature 453}, 7197 (6 2008), 869--878.

\bibitem{lancetadhd}
{\sc Lou, H.~C., Henriksen, L., and Bruhn, P.}
\newblock Focal cerebral dysfunction in developmental learning disabilities.
\newblock {\em The Lancet 335}, 8680 (1 1990), 8--11.

\bibitem{drmackay}
{\sc MacKay, D.~J.~C.}
\newblock {\em Information Theory, Inference, and Learning Algorithms}
\newblock Cambridge University Press, 2003

\bibitem{dadmackay1}
{\sc MacKay, D.~M.}
\newblock A high-speed electronic function generator
\newblock {\em Nature 159}, 4038 (3 1947), 406--407

\bibitem{mann}
{\sc Mannfolk, P., Wirestam, R., Nilsson, M., Stahlberg, F., and Olsrud, J.}
\newblock Dimensionality reduction of fMRI time series data using locally
  linear embedding.
\newblock {\em MAGMA}, 23(5-6):327--338, Dec 2010.

\bibitem{melnikov}
{\sc Mel'nikov, M.~S.}
\newblock Analytic capacity: discrete approach and curvature of measure
\newblock {\em Sbornik: Mathematics 186}, (6):827-846, 1995

\bibitem{sol1}
{\sc Mill, J.~S.}
\newblock {\em A System of Logic, Ratiocinative and Inductive: Being a Connected View of the Principles of Evidence and the Methods of Scientific Investigation, Volume I}.
\newblock John W. Parker, 1843.


\bibitem{mink}
{\sc Minkowski, H.}
\newblock Raum und Zeit.
\newblock {\em Physikalische Zeitschrift} 10 (1908), 75--88.

\bibitem{newton}
{\sc Newton, I.}
\newblock {\em Philosophiae naturalis principia mathematica}.
\newblock J. Societatis Regiae ac Typis J. Streater, 1687.

\bibitem{ogawa3}
{\sc Ogawa, S., Lee, T.~M., Kay, A.~R., and Tank, D.~W.}
\newblock Brain magnetic resonance imaging with contrast dependent on blood
  oxygenation.
\newblock {\em Proceedings of the National Academy of Sciences 87}, 24 (1990),
  9868--9872.

\bibitem{pinv}
{\sc Penrose, R.}
\newblock A generalized inverse for matrices.
\newblock {\em Mathematical Proceedings of the Cambridge Philosophical Society}, 51:406--413, 7 1955.


\bibitem{neuronuclear}
{\sc Popham, M.~G.}; {\sc Potchen, E.~J.,  McCready, V.~R.} (Eds.)
\newblock Numerical Methods for the Detection of Abnormalities in Radionuclide Brain Scans.
\newblock {\em Progress in Nuclear Medicine 1: Neuro-Nuclear Medicine}.
\newblock University Park Press, 1972. 117-141

%

\bibitem{rayleigh}
{\sc Rayleigh, J.~W.~S.}
\newblock {\em The theory of sound}.
\newblock MacMillan and Co. London, 1877.

\bibitem{psychassess}
{\sc Robins, L.~N., and Helzer, J.~E.}
\newblock Diagnosis and clinical assessment: The current state of psychiatric
  diagnosis.
\newblock {\em Annual Review of Psychology 37}, 1 (1986), 409 -- 432.

\bibitem{lle1}
{\sc Roweis, S.~T., and Saul, L.~K.}
\newblock Nonlinear dimensionality reduction by locally linear embedding.
\newblock {\em Science 290}, 5500 (2000), 2323--2326.

\bibitem{roysher}
{\sc Roy, C.~S., and Sherrington, C.~S.}
\newblock On the regulation of the blood-supply of the brain.
\newblock {\em Journal of Physiology 11}, 1-2 (Jan 1890), 85--158.


\bibitem{LLE2}
{\sc Saul, L.~K., and Roweis, S.~T.}
\newblock Think globally, fit locally: unsupervised learning of low dimensional
  manifolds.
\newblock {\em Journal of Machine Learning Research 4\/} (Dec. 2003), 119--155.

\bibitem{seunglee}
{\sc Seung, H.~S., and Lee, D.~D.}
\newblock The manifold ways of perception.
\newblock {\em Science 290}, 5500 (2000), 2268--2269.

\bibitem{schmidt}
{\sc Schmidt, E.}
\newblock \"Uber die Aufl\"osung linearer Gleichungen mit Unendlich vielen unbekannten
\newblock {\em Rendiconti del Circolo Matematico di Palermo 25} 1 (1908), 53--77.

 
\bibitem{sherr}
{\sc Sherrington, C.~S.}
\newblock {\em Man on His Nature}.
\newblock Gifford lectures, Edinburgh. University Press, 1951.

\bibitem{memoryreview}
{\sc Simons, J.~S., and Spiers, H.~J.}
\newblock Prefrontal and medial temporal lobe interactions in long-term memory.
\newblock {\em Nature Reviews Neuroscience 4}, 8 (08 2003), 637--648.

\bibitem{sternberg}
{\sc Sternberg, S.}
\newblock High-speed scanning in human memory.
\newblock {\em Science 153}, 3736 (1966), 652--654


\bibitem{sylvester}
{\sc Sylvester, J.~J.}
\newblock A demonstration of the theorem that every homogeneous quadratic polynomial is reducible by real orthogonal substitutions to the form of a sum of positive and negative squares
\newblock {\em Philosophical Magazine Series 4 (4)} 23 (1852), 138--142

\bibitem{sfs}
{\sc Whitney, A.}
\newblock A direct method of nonparametric measurement selection.
\newblock {\em Computers, IEEE Transactions on C-20}, 9 (sept. 1971), 1100 --
  1103.

\bibitem{scid}
{\sc Williams, J., Gibbon, M., First, M., Spitzer, R., Davies, M., Borus, J.,
  Howes, M., Kane, J., Pope, H., and Rounsaville, B.}
\newblock The structured clinical interview for dsm-iii-r (scid). ii. multisite test-retest reliability.
\newblock {\em Archives of general psychiatry 49}, 8 (08 1992).

\bibitem{mlle}
{\sc Zhang, Z. and Wang, J.}
\newblock MLLE: Modified Locally Linear Embedding Using Multiple Weights
\newblock {\em Advances in Neural Information Processing Systems} 19 (2007): 1593


\bibitem{autismpsych}
{\sc Zilbovicius, M., Boddaert, N., Belin, P., Poline, J., Remy, P., Mangin,
  J., Thivard, L., Barth{\'e}l{\'e}my, C., and Samson, Y.}
\newblock Temporal lobe dysfunction in childhood autism: a PET study.
\newblock {\em American Journal of Psychiatry 157}, 12 (2000), 1988--1993.


\end{thebibliography}
\end{document}